\documentclass[12pt]{article}
\usepackage{amsmath}
\usepackage{times}
\usepackage{graphicx}
\usepackage{color}
\usepackage{multirow}
\usepackage[authoryear]{natbib}
\usepackage{rotating}
\usepackage{bbm}
\usepackage{latexsym}
\usepackage{ccaption}
\newcommand{\captionfonts}{\normalsize}

\makeatletter  
\long\def\@makecaption#1#2{%
  \vskip\abovecaptionskip
  \sbox\@tempboxa{{\captionfonts #1: #2}}%
  \ifdim \wd\@tempboxa >\hsize
    {\captionfonts #1: #2\par}
  \else
    \hbox to\hsize{\hfil\box\@tempboxa\hfil}%
  \fi
  \vskip\belowcaptionskip}
\makeatother   

\begin{document}
\hspace{13.9cm}1

\ \vspace{20mm}\\

{\LARGE Gain modulation of probabilistic selection without synaptic relearning}

\ \\
{\bf \large Elif K\"oksal-Ers\"oz$^{\displaystyle 1, 2}$, Pascal Chossat$^{\displaystyle 3, \displaystyle 4}$, Fr\'ed\'eric Lavigne$^{\displaystyle 3, \displaystyle 5}$}\\
{$^{\displaystyle 1}$Inria, Villeurbanne, France}; 
{$^{\displaystyle 2}$Cophy Team, Lyon Neuroscience Research Center, Bron, France};
{$^{\displaystyle 3}$MathNeuro Project-Team, Inria, Montpellier, France}, 
{$^{\displaystyle 4}$Laboratoire Jean-Alexandre Dieudonn\'e , Universit\'e C\^ote d’Azur, Nice, France};
{$^{\displaystyle 5}$BCL, Universit\'e C\^ote d’Azur, CNRS, Nice, France.}


\thispagestyle{empty}
\markboth{}{NC instructions}
\ \vspace{-12mm}\\
%
\begin{center} {\bf Abstract} \end{center}
Adaptation of behavior requires the brain to change goals in a changing environment. Synaptic learning has demonstrated its effectiveness in changing the probability of selecting actions based on their outcome. In the extreme case, it is vital not to repeat an action to a given goal that led to harmful punishment. The present model proposes a simple neural mechanism of gain modulation that makes possible immediate changes in the probability of selecting a goal after punishment of variable intensity. Results show how gain modulation determine the type of elementary navigation process within the state space of a network of neuronal populations of excitatory neurons regulated by inhibition. Immediately after punishment, the system can avoid the punished populations by going back or by jumping to unpunished populations. This does not require particular credit assignment at the `choice' population but only gain modulation of neurons active at the time of punishment. Gain modulation does not require statistical relearning that may lead to further errors, but can encode memories of past experiences without modification of synaptic efficacies. Therefore, gain modulation can complements synaptic plasticity.

\section{Introduction}\label{sec1}

Adaptation of behavior requires the brain to select actions that bring benefits and to avoid those that bring costs. Selection of the most probably rewarded action requires updating the relation between actions and their rewards and punishments from past experience \citep{rescorla1971variation, sutton2018reinforcement, behrens2007learning}. In a stable environment, the probability of the outcomes is best estimated by experiences going back a long way to ensure exploitation of the rewarded actions. If the probabilistic structure of feedback changes, actions become prone to errors due to uncertainty on expected feedback. Exploration of different actions and relearning must then update the action-outcome probability \citep{cohen2007should, domenech2020neural} through statistical learning of trials and errors \citep{Lazartigues2021, rey2022learning, lazartigues2023probability}. This raises the question of the degree of recency of the experiences and of the intensity of the outcomes to be taken into account. 

Animal studies report that ancient and recent rewards are both memorized \citep{corrado2005linear, fusi2007neural, bernacchia2011reservoir}. Changes in the rate of rewards and punishments can be adjusted by changes in a single learning rate to update synaptic efficacies \citep{behrens2007learning, fusi2007neural, nassar2010approximately, nassar2012rational}. Computational models have investigated a learning rate at synapses connecting a context to different actions that depends on the magnitude of the error signal to optimize the weighting of old and recent experiences over multiple timescales \citep{iigaya2016adaptive, iigaya2019deviation}.

When changes in the environment are transient, different structures of action-feedback are associated to transiently alternating contexts. In this case, it is beneficial to adapt behaviors to the transient state without forgetting the previous state and hence without the need to relearn it. Synaptic relearning leads to forgetting of the previous probabilistic structure of action-outcome relations. The previously learned and forgotten environment has then to be learned again through repetition of trials and errors, even though at a fast rate. However, even more dramatically, in the extreme case of severe and dangerous punishment, the action that led to it should not be repeated. Some errors must not be made twice. This does not give time (or oppotunity) for statistical relearning. Then is it possible to change actions without synaptic relearning? And, in the extreme case, without the need for any further learning trial? 

Various cortical functions are reported to rely on the modulation of the input-output gain at the level of neurons, defined as the slope of the neurons transfer function \citep{salinas2000gain, carandini2012normalization, eldar2013effects, ferguson2020mechanisms, debanne2019plasticity}. At the level of network behavior, gain modulation changes large scale network dynamics \cite{shine2021computational} and the correlation of neuronal output activity \citep{doiron2016mechanics}, giving the network the computational ability to change which neurons are activated by a context even though values of synaptic efficacy are fixed \citep{Lavigne2008}. Further, a recent computational model has reported that a network can switch activation between sequences of neurons coding for items depending on the modulation of the gain of these neurons \citep{koksal_ersoz_dynamic_2022}. Gain modulation was shown as efficient in changing the probability of activating an item or another in the network state space for a fixed synaptic matrix. Here we investigate the conditions under which neuronal gain alone enables switching from one action to another without the need for synaptic relearning, and possibly with immediate effect in case of punishment. This raises the question of the assignment of punishment to the sequence of neurons that led to the punished behavior (\textit{e.g.,}\cite{li2020learning, asaad2017prefrontal, he2015distinct, friedrich2011spatio, liu2021cell}).

\section{Methods \label{sec:methods}}

The model has been directly inspired by \citep{koksal_ersoz_dynamic_2022}, where the retrieving of multiple sequences in a collection of $P$ learned states $\xi^1,\cdots,\xi^P$ has been investigated using the framework introduced in \citep{aguilar_latching_2017, koksal_ersoz_neuronal_2020}. In this model each learned state is a dynamically stable `pattern' made of two active units coding for populations of neurons in the neural network, the others units being inactive. Moreover, these patterns can be destabilized under the effect of short term synaptic depression (STD), hence allowing for dynamics of activation of patterns in the network state space.

In \citep{aguilar_latching_2017, koksal_ersoz_neuronal_2020} any two consecutive learned patterns share one unit, so that the patterns form a chain of overlapping states $\xi^1 - \xi^2 -\cdots - \xi^P$, where $P$ is the number of patterns $\xi^m$. It has been shown that, under the effect of noise, the overlapping condition allows to produce a sequential stochastic dynamics, one state `jumping' to the next with high probability. This behavior is called latching dynamics.

In \citep{koksal_ersoz_dynamic_2022} we analyzed the case when, as the system starting from $\xi^1$ reaches a given state $\xi^m$ in the chain (the branching `node'), there is a choice among several continuing branches. In the simplest case, which is considered in the present work, the chain splits at the node into two chains $\xi^m, \xi^{m+1} \cdots,\xi^q$ (branch 1, Br-1) and $\xi^m, \xi^{q+1}, \cdots,\xi^{P}$ (branch 2, Br-2).

As shown in \citep{koksal_ersoz_dynamic_2022}, as long as the connectivity matrix is symmetric, the probabilities that the dynamics starting from $\xi^1$ continues on either branch 1 or branch 2 are equal, but whenever the weights of connections from $\xi^m$ to $\xi^{m+1}$ are greater that those from $\xi^m$ to $\xi^{q+1}$, the probability to continue on branch 1 is larger than the probability to continue on branch 2. The branching network which we numerically investigated is pictured in Figure \ref{fig1}A. Its parameters are $N=10$, $P=9$, $m=4$ and $q=7$.  

The equations for the units are derived from \citep{amari}, in which we have replaced the membrane potential $u_i$ of each unit $i$ by the activity $x_i=S(u_i)=1/(1+e^{-\gamma u_i})$. The variables are now the activities which take values in the interval $[0,1]$ after $S^{-1}(x_i)$ has been replaced by its polynomial expansion (we chose the simplest, linear approximation). The inhibition effects within the network are modeled by a term proportional to the averaged activity \citep{lerner2012}.

The equation for unit $i$ reads
\begin{equation}
\dot{x_i} = x_i(1-x_i)\left(-\frac{4}{\gamma} x_i + \sum_{j=1}^N J_{i,j} x_j - \lambda\sum_{j=1}^N x_j - \lambda\nu_ix_i\right) + \eta
\label{eq:x}
\end{equation}
where $\gamma$ is the gain and $\lambda$ is the inhibitory coefficient. The coefficient $\nu_i$ accounts for the possibility of short-range inhibitory loops between excitatory neurons and inhibitory interneurons. This allows for selective self-inhibition of excitatory neurons, which depends on the number of afferent connections to excited units. In our case we take $\nu_i=0$ for $i\neq m+1$ and $\nu_{m+1}=1$ (unit $m+1$ receives inputs from units $m, m+1$ and $q+1$). Finally $\eta$ is a noise term. We refer the reader to \citep{koksal_ersoz_dynamic_2022} for details. Note that $x_i=0$ or $1$ are always solutions of \eqref{eq:x}. Hence, any state such that $x_i=0$ or $1$, $i=1\dots,N$ is a steady-state of the system. The learned states are stable solutions of this type.

The STD is expressed as follows. Let $J_{ij}(t)$ be the strength of the connection from units $j$ to $i$ at time $t$. We write $J_{ij}(t)=J_{ij}^{max} s_j(t)$, where $J_{ij}^{max}$ is the connectivity matrix resulting from the learning process and $s_j(t)$ follow the STD law given in \citep{tsodyks1997}, which is equivalent to
\begin{equation}
\tau_r \dot{s_i} = 1 - s_i - \rho s_i x_i.
\label{eq:s}  
\end{equation}
The weights $J_{ij}^{max}$ are computed from the simple Hebbian rule $J_{ij}^{max}=\sum_k\xi^k_i\xi^k_j$.


The parameter values of \eqref{eq:x}-\eqref{eq:s} are optimized from \cite{koksal_ersoz_dynamic_2022} as $\lambda = 0.6, I = 0, \nu_4 = 1, \rho = 300, \tau_r = 1.2$, $\gamma = 10$ without punishment, $\gamma = \{9, 5, 3.3, 2.5\}$ with punishment (from weak (10\%) to strong (75\%) rates). 

\section{Results and Discussion \label{sec:results}}

The model presented here allows mathematical analysis and simulation of the effect of gain modulation on the selection of goals towards which to direct actions. We investigate a system of $N=10$ units ($x_i$, for $i \in \{1, … 10 \}$) encoding 9 patterns (states) (Fig \ref{fig1}A), which we write (A,B, \dots, I) for convenience (Fig \ref{fig1}B). The units are placed in a 3-node graph where units 1-3 are along branch 0, units 5-7 along branch 1, units 8-10 along branch 2 and unit $x_4$ is the branching node connected to $x_3$, $x_5$ and $x_8$ (Fig \ref{fig1}A). Such network architecture could correspond to sequences of internal representations of successive goals to go. The branching synaptic architecture correspond to a classical Y-maze \citep{ghafarimoghadam2022review} for which the network codes successive goals to the final goal at the end of branches, which can be rewarded or punished. The activation of two units side by side (\textit{e.g.}, $x_2$ and $x_3$) code for a pattern corresponding to a network state (\textit{e.g.}, B). A change in pattern activated would correspond to the activation of a new goal where to orient actions in the Y-maze. For the sake of clarity, We focus here on the elementary building block of network behavior at a single branching node \citep{koksal_ersoz_dynamic_2022}. But the results presented here can be generalized to more complex networks involving 4-ways or more branchings and \textit{branches in branches}). 

\begin{figure}
    \centering
    \includegraphics[scale = 1]{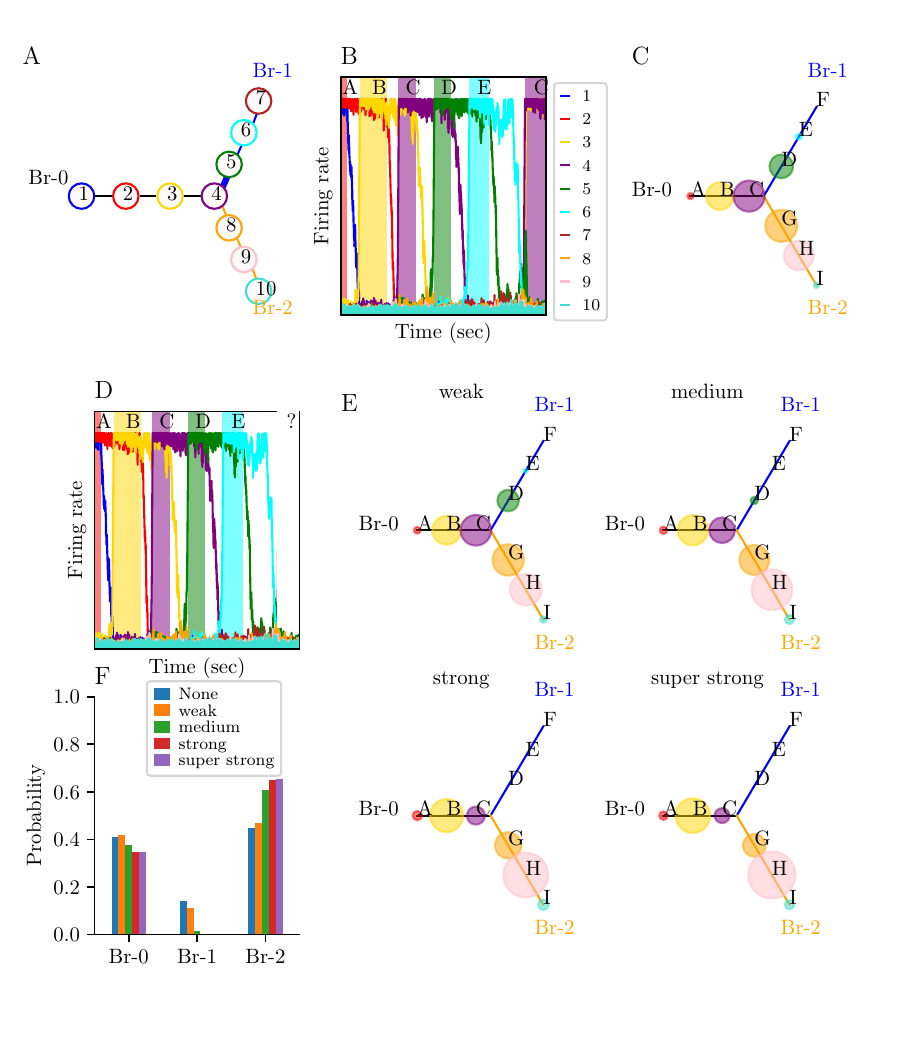}   
    \caption{Branching behavior of a $N = 10$ units network at Trial $T$. (A) Network architecture of 10 units represented by numbered circles. Unit 4 is a branching node between three branches 0, 1 and 2. The synaptic efficacy between units 4 and 5 is 10\% stronger than between units 4 and 8. (B) Network behavior is described from the activation of units 1 and 2 in the initial branch (Br-0). In this example, activation propagates from units 1 to 4 in a sequence, and \textbf{(caption continues next page)}}
    \label{fig1}
\end{figure}
\begin{figure}[th!]
  \contcaption{continues to units 5 and 6 in branch 1 (Br-1), thanks to the stronger efficacy between units 4 and 5. Line colors correspond to the activation of units and colored areas correspond to patterns coded by these units. (C) The network embeds 9 patterns coded by two connected units which follow each other in a sequence (letters A to I). Unit 4 is part of patterns C, D, and G, themselves parts of branches 0, 1 and 2, respectively. Circles size is proportional to the probability of activation of the patterns. In this same example as in (B), the sequential activation propagates from pattern A to pattern E in Br-1, again thanks to the stronger efficacy between units 4 and 5 of pattern D. Pattern D is the last possible pattern to activate in Br-1. After reaching it, the network randomly jumps to another pattern (here pattern C). (D) If punishment is given at the activation of pattern E, the network behavior after this pattern depends on the punishment rate. In this example, the network stops without activating any pattern. (E) Probability of activation of the patterns in the 3 branches immediately after punishment during the punished trial $T$, as a function of the level of punishment. Circles size is proportional to the probability of activation of the patterns immediately after punishment. (F) Probability of activation of any pattern in each of the 3 branches for all levels of punishment (1000 simulations for each level). When the punishment is weak (10\% decrease in neuronal gain), the system can still activate patterns D or E but in only in 11\% of the trials. For a medium punishment (50\% decrease in gain), this ratio decreases to 1\%. Whereas for a strong punishment (66\% decrease in gain), the system does not activate patterns along Br-1 anymore. Instead, the system activates patterns along branch 2 (Br-2) with a 40\% increase from weak to strong punishment.}
\end{figure}

\newpage

The synaptic coupling coefficient between units 4 and 5 (branch 1) was 10\% stronger than between units 3 and 4 (branch 0), and units 4 and 8 (branch 2). We study if and how gain modulation can change the sequence of neurons activated during a punished trial and on the immediately following trial. For simulations, the system was initialized at pattern A and punishment was applied to units 5 and 6 (coding for pattern E) when they became activated during the trial $T$. Punishment was assigned to neurons active at the time of feedback \citep{asaad2017prefrontal}, assumed to depend on punishment signaling (noradrenaline, serotonin and/or dopamine; \citep{tanaka2009serotonin,cohen2012neuron, oleson2012subsecond, cohen2015serotonergic, michely2022serotonin} that decreases neuronal gain \citep{henze2000dopamine, Bandyopadhyay2007dopaminergic, noudoost2011control, seillier2017serotonin}.

\begin{figure}
    \centering
    \includegraphics[scale = 1]{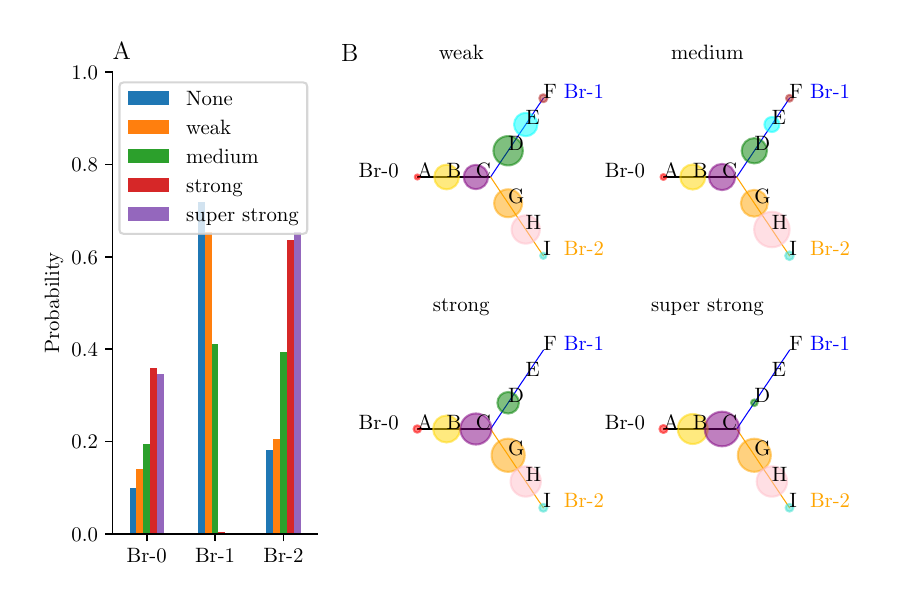}
    \caption{Branching behavior of a $N = 10$ units network (see Figure \ref{fig1}A-C) at Trial $T+1$. (A) Probability of activation of patterns in the 3 branches during the punished trial $T$, for all levels of punishment (1000 simulations for each level). In the absence of any punishment (`None'), the system takes branch 2 (Br-2) in 71\% of the trials. For a weak punishment, the system activates branch 1 (Br-1) in 65\% of trials and Br-1 in 20\% of trials. Moderate punishment equalizes the probability between the two branches (41\% for Br-1 and 39\% for Br-2). For strong punishment, the networks activates the Br-2 only (0\% for Br-1 vs 64\% for Br-2). (B) Probability of activation of the patterns in the 3 branches at trial $T+1$ as a function of the level of punishment. Circles size is proportional to the probability of activation of the patterns. The effect of punishment at trial $T$ is maintained in the trial $T+1$. Strong punishment at trial prevents from activating the punished branch. Activation either goes back to Br-0 or jumps to Br-2.}
    \label{fig2}
\end{figure}

Results show that punishment-dependent gain modulation changes the probability of activating the punished and unpunished patterns. Further, the effects of punishment observed immediately after punishment \textit{during} the punished trial $T$ (Fig \ref{fig1}E, \ref{fig1}F) is maintained at the following trial $T+1$ (Fig \ref{fig2}A, \ref{fig2}B).

Three main network behaviors are observed depending on the level of punishment:

\begin{itemize}
\item 
In the absence of punishment, the gain is the same in all units in the two branches. In that case, the stronger synaptic connection between the branching units 4 and unit 5 drives the network behavior. It induces more frequent activation of branch 1 by activating patterns D then E after the initial sequence A-B-C (Figure \ref{fig1}B). The model reproduces an exploitation strategy that increases the probability of reward \citep{cohen2007should, domenech2020neural}. 

\item 
After punishment of medium intensity, the gain is decreased in the units 5 and 6 (coding for pattern E) active at time of punishment. This decreases the probability of recalling patterns D or E in the punished branch 1. This occurs immediately during the punished trial $T$ (Fig \ref{fig1}E) and in the following trial $T+1$ (Fig \ref{fig2}B). The lower gain in the punished branch 1 makes neurons populations less responsive to input activity coming from the initial branch 0 (patterns A-B-C). Given the stronger synaptic efficacy between the punished branch 1 and the branching node 4, the system still activates the punished branch 1 but with lower probability (Fig \ref{fig2}A). The balanced probabilities to select the two branches corresponds to an exploration strategy to search for the most rewarded or less punished goals \citep{cohen2007should, domenech2020neural}. 

\item 
After strong punishment, the gain is strongly decreased in the punished units 5 and 6. After the first regular sequence A-B-C (Fig \ref{fig2}A), the activation of patterns along branch 1 is stopped immediately at trial $T$ (Fig \ref{fig1}F) and is avoided at trial $T+1$ (Fig \ref{fig2}B), although this branch 1 is the most  strongly (synaptically) associated to the branching node. In that case, the network can go back or switch directly to branch 2. The model reproduces an avoidance strategy that prevents from further strong punishment. 

\end{itemize}

\begin{figure}
    \centering
    \includegraphics[scale = 1]{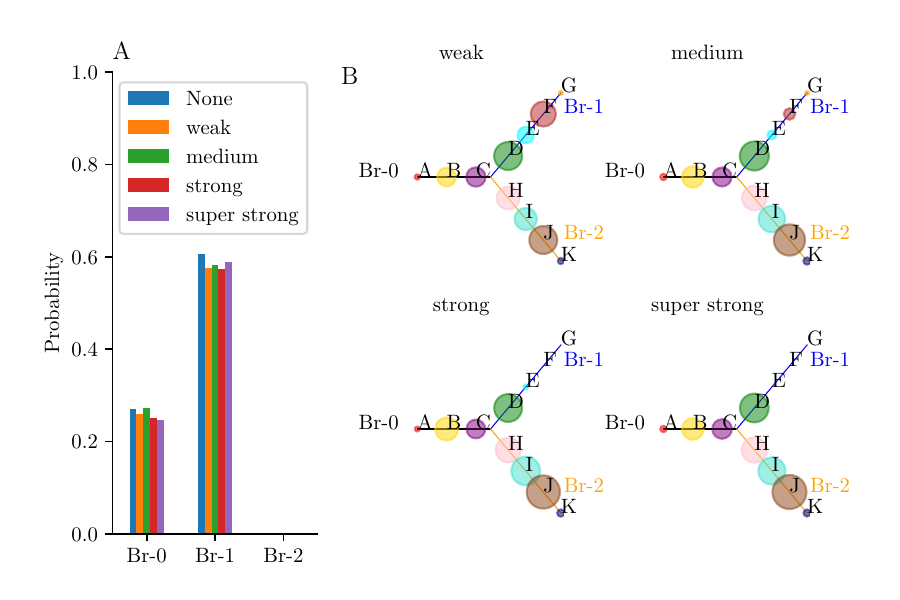}
    \caption{Branching behavior of a $N = 12$ units network at Trial $T+1$. (A) Probability of activation of patterns in the 3 branches immediately after punishment \textit{during} the punished trial $T$, for all levels of punishment (1000 simulations for each level). Punishment does not prevent from activating pattern D in branch 1 (Br-1) at trial $T$, due to the stronger synaptic connection with the branching node 4. (B) Probability of activation of the patterns in the 3 branches at trial $T+1$ \textit{following} the punished trial $T$, as a function of the level of punishment. Circles size is proportional to the probability of activation of the patterns. Increased punishment decreases the probability of activation of patterns E and F in Br-1 at trial $T+1$. Strong punishment at trial $T$ even prevents from activating the punished branch again at trial $T+1$. The network activates pattern D at the beginning of Br-1, but then either goes back to branch 0 (Br-0) or jumps to branch 2 (Br-2).}
    \label{fig3}
\end{figure}

In the model presented in Figures \ref{fig1} and \ref{fig2}, punishment decreased the gain of units (5-6) directly connected to the branching unit (4) and impacted the choice behavior at unit 4 without exploration of the branch1. However, sequences of activation of units preceding a feedback can be longer, leading to punishment of distant units not directly connected to the branching unit 4 (\textit{e.g.}, 6-7). In that case the system does not `know' which branch was punished when arriving at the branching unit (because unit 5 connected to unit 4 was not punished). This scenario was tested in a network with longer branches (Fig \ref{fig3}).

When punishment arrived at patterns distant from the branching node (Fig \ref{fig3}A), gain was decreased on units 6 and 7 and remained unchanged in all other units of the punished branch that were not active at the time of punishment (here unit 5). Results show that such distant assignment of punishment is sufficient to switch branch at trials $T$ and $T+1$ (Fig \ref{fig3}B). The system can still activate the unpunished unit 5 at the beginning of the punished branch, but switches branch before arriving at the punished units (6-7). Then the network can begin in the punished branch but switches to and fully activates the other (unpunished) branch. This navigation process makes possible to switch behavior after punishment. It is not exclusive of synaptic eligibility traces \cite{he2015distinct} but does not need assignment of feedback to the whole sequence of units that were activated before the punishment. This way the punishment assignment obeyed a simple mechanism combining punishment with only units active at the time of feedback. Such simple mechanism of assignment is made efficient even in long sequences of patterns, thanks to the network's ability to navigate forwards, backwards or jump branches in its state space, depending on the gain of the neuron population. 

Interestingly, different navigation processes can be generated by different types of sequential activation in the phase space of the network. The punished state can be avoided by an activity that can either 
\begin{itemize}
    \item[-] go back to the starting branch 0 coding for the context,
    \item[-] stay at the pattern preceding the punished pattern in branch 1,
    \item[-] jump directly to the unpunished branch 2.
\end{itemize}

The model exhibits elementary navigation processes within the network's state space ({\em building blocks} \citep{koksal_ersoz_dynamic_2022}) that depend on neuronal gain. Such processes allow the network to adapt its behavioral strategy to synaptically learned reward and to the level of punishment. Taken as a whole, results show that gain modulation switches the network behavior between exploitation and exploration behaviors: 
\begin{itemize}
    \item[-] an approach exploitation behavior: a synaptically driven activation of the goal learned as rewarded (branch 1) rather than the goal learned as punished (branch 2) (\ref{fig2}A, no punishment),
    \item[-] an exploration behavior: a balanced selection between the goal synaptically learned as rewarded and that has been recently weakly punished (branch 1) and the goal synaptically learned as punished (branch 2) (Fig \ref{fig2}B and Fig \ref{fig3}B, weak punishment), 
    \item[-] an avoidance exploitation behavior: a gain driven blocking of the strongly punished branch 1 and switch to the learned as less punished branch 2 (Fig \ref{fig2}B and Fig \ref{fig3}B, strong punishment). 

\end{itemize}

The avoidance strategy after strong punishment allows to not repeat harmful errors during trial and error learning. We hope that these can provide a framework for modeling and experimental approaches investigating the effects of punishment on gain modulation and goal selection without synaptic relearning.

\section*{Conclusion}
From a learning point of view, the present results indicates that neuronal gain can embed knowledge on the relation between goals and outcomes. This way the value of neuronal gain contributes to store memories of past experiences \citep{zhang2003other} without changes in the synaptic matrix. Such gain-based neural learning could complete synaptic learning in the alternation between exploitation, exploration and avoiding strategies. Synaptic learning allows knowledge to change rapidly and/or slowly depending on the volatility of the environment \citep{iigaya2019deviation}. Given that gain modulation alters neuronal excitability, it could provide an alternative means of storing knowledge at the microscopic neuronal level, in addition to synapses \citep{daoudal2003long, abraham2019plasticity, debanne2019plasticity}. The present model shows that local gain decrease by punishment signaling \citep{henze2000dopamine, Bandyopadhyay2007dopaminergic, noudoost2011control, seillier2017serotonin} is able to change network behavior at the macroscopic level without synaptic relearning. In the framework of the effects of gain on the macroscopic network behavior, intrinsic gain learning raises a number of points for further study: to what extent the effects of gain on network behavior differ from those of synaptic learning? under which conditions of intensity and frequency reward or punishment affect transient and/or long-term behavior? and what are the interactions between the variations in gain and synaptic efficacy when both change at the same time? Further, in the present model, patterns learned in the synaptic matrix are assumed to code for goals that orient actions. Model refinements could couple this network with a network whose patterns code for the sequences of actions between the different goals.

In its current state, the present models shows that gain-based neuronal learning enables a modulation of knowledge activated in memory in a less synapse-dependent way though not altering knowledge previously stored in the synapses. In the case of punishment, gain-based learning could then give the system the necessary time for synaptic relearning without repeating harmful errors.

\section*{Acknowledgements}
The authors gratefully thank Pascal Benquet and Gianluigi Mongillo for insightful discussions on a previous version of this article. This work was supported by the Hebbian ANR-project (ANR-23-CE28-0008).

\bibliographystyle{APA}

\end{document}